# Observation and enhancement of room temperature bilinear magnetoelectric resistance in sputtered topological semimetal Pt$_3$Sn

Yihong Fan[1], Zach Cresswell[3], Yifei Yang[1], Wei Jiang[1], Yang Lv[1], Thomas J. Peterson[2], Delin Zhang[1], Jinming Liu[1], Tony Low[1], and Jian-Ping Wang[1]*

[1]*Department of Electrical and Computer Engineering, University of Minnesota, 200 Union St. SE, Minneapolis, Minnesota 55455, USA*

[2]*School of Physics and Astronomy, University of Minnesota, 116 Church St. SE, Minneapolis, MN 55455, USA*

[3]*Department of Chemical Engineering and Materials Science, University of Minnesota 421 Washington Ave. SE, Minneapolis, MN 55455, USA*

*Corresponding author: jpwang@umn.edu

**Abstract**

Topological semimetal materials have become a research hotspot due to their intrinsic strong spin-orbit coupling which leads to large charge-to-spin conversion efficiency and novel transport behaviors. In this work, we have observed a bilinear magnetoelectric resistance (BMER) of up to 0.0067 nm$^2$A$^{-1}$Oe$^{-1}$ in a singlelayer of sputtered semimetal Pt$_3$Sn at room temperature. Different from previous observations, the value of BMER in sputtered Pt$_3$Sn does not change out-of-plane due to the polycrystalline nature of Pt$_3$Sn. The observation of BMER provides strong evidence of the existence of spin-momentum locking in the sputtered polycrystalline Pt$_3$Sn. By adding an adjacent CoFeB magnetic layer, the BMER value of this bilayer system is doubled compared to the single Pt3Sn layer. This work broadens the material system in BMER study, which paves the way for the characterization of topological states and applications for spin memory and logic devices.

Improving charge-to-spin conversion efficiency has been a research hotspot in spintronics ever since the discovery of spin-orbit torque [1-12]. Topological materials, including topological insulators [5-7] and Dirac and Weyl semimetals [8-12] (DSM and WSM) are proposed to have a large charge-to-spin conversion ratio due to the existence of surface or edge states and strong spin-orbit coupling. Large spin Hall angle and small switching current density have been reported in topological insulators at room temperature [5,6], which shows great application potential in energy-efficient spin-orbit torque devices. Unlike topological insulators though, bulk topological states exist in semimetals, where the conduction and valence bands are connected by nodes and lines in the bulk. The contribution of the bulk to conduction makes semimetals more favorable for industrial spin memory applications compared to topological insulator materials. Relatively large values of spin Hall angle have been reported in several WSM and DSM systems [8-12].

Proving the existence of topological states has become the key factor in such research. Different methods are used to characterize the existence of topological states, the most common method being angle-resolved photoemission spectroscopy (ARPES). ARPES gives direct evidence of the topological features such as the Weyl nodes [13-15], but requires a single crystal sample and in-situ measurement. Another way is to use negative longitudinal magnetoresistance (NLMR) as evidence of topological states [16-21], which may contain a variety of transport contributions [19-21] and is not universal for all topological materials. An easier and universal method to show the evidence of strong spin-orbit coupling and topological states is highly desirable. Spin-momentum locking (SML) [22-24], which fixes the spin direction of the electrons perpendicular to the momentum direction, is strong evidence of the existence of topological features, including the Dirac cone [22,23] and Weyl nodes [24], in the band structure. However, direct detection of SML requires a non-local detection [22], which necessitates a different device structure and fabrication method compared to other transport measurements such as spin-orbit torque (SOT) switching and magnetoresistance measurements. Moreover, since SML occurs along certain crystalline directions, direct detection of SML may not be suitable for polycrystalline materials. Recently, the discovery of bilinear magnetoelectric resistance (BMER) provides an easy way to prove the existence of spin-momentum locking, and as a result, the topological states [25-29].

The illustration of BMER is shown in Figure 1. With an electric field $E$ applied to the sample, a second-order variation of the electron distribution exists, which leads to a spin current that is proportional to $E^2$ [25], as shown in Figure 1(a). With a magnetic field applied, the population of the favored spin changes, which partially converts the spin current to a charge current, resulting in high and low resistance states, as shown in Figure 1(b) and (c). Existing in materials with topological spin texture, BMER changes linearly with both the magnetic and electrical fields without the presence of a magnetic layer. With a similar measurement setup to unidirectional spin Hall magnetoresistance (USMR) [30-34] and unidirectional magnetoresistance (UMR)[35], BMER can be measured with a simple two-terminal bar structure. This makes BMER a prospective candidate to indicate the existence of topological features in transport measurements. Theoretically, BMER is observable even if the topological material is polycrystalline, which makes BMER even more favorable for industrial applications. However, up to now, no such experiments have been carried out.

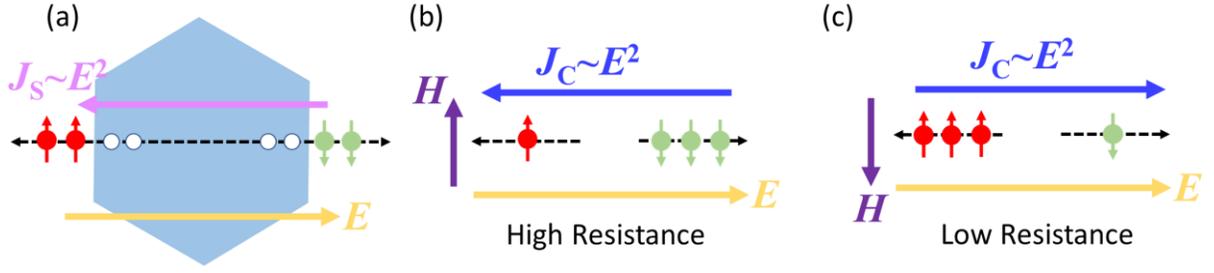

Figure 1. Illustration of BMER. (a) The electric field $E$ leads to a second-order change of the electron distribution, which results in a spin current proportional to $E^2$. (b) and (c) With an applied magnetic field, electron populations of different spins change, which result in high and low resistance states.

Binary Pt-Sn alloys have been explored both theoretically and experimentally [36-38] as a promising topological material. Different phases and compositions, such as $PtSn_4$ and $Pt_3Sn$, are proposed and demonstrated to have different topological properties [36-38]. With both topological surface states and bulk conduction, sputtered $Pt_3Sn$ was proved to be a topological insulator at the surface and type II Weyl semimetal in the bulk, with a large spin Hall angle as well as a robust NLMR [38]. Its small resistivity compared to other topological materials such as $Bi_2Se_3$[6], as well as the easy fabrication process by sputtering [38], makes $Pt_3Sn$ an industrially-compatible candidate for SOT-based memory technologies.

In this work, we report that the BMER value reaches 0.0067 nm$^2$A$^{-1}$Oe$^{-1}$ in sputtered Dirac semimetal $Pt_3Sn$ polycrystalline single layer at room temperature. The BMER is observed in plane and absent in the YZ plane, suggesting the polycrystalline feature cancels the YZ plane signal. Additionally, the magnitude of the BMER value is enhanced by ~2 times with an adjacent ferromagnetic CoFeB layer, which shows the potential for interfacial manipulation of the topological structures. The existence and manipulation of BMER in polycrystalline $Pt_3Sn$ film has paved the way for the design and characterization of the topological features in polycrystalline materials.

A single layer of $Pt_3Sn$ (15nm) is deposited on a single-crystal MgO substrate with an eight-target sputtering system under base pressure $5\times10^{-8}$ Torr. The sample was co-sputtered from targets of Pt and $PtSn_4$ and the substrate was baked, deposited, and post-annealed for 1 hour at 350˚C. A 4nm Al layer is deposited on the top to protect the surface from oxidization. The Al capping layer is fully oxidized and does not contribute to conduction. X-ray diffraction (XRD) measurements were then taken of the film in a Bruker D8 Discover diffractometer, utilizing a 2-D detector to observe the out-of-plane texturing, and a Co-Kα1 source.

The XRD results are shown in Figure 2, where the inner image shows the pattern on the 2-D detector. Diffraction peaks for [111] and [002] direction $Pt_3Sn$ are present. The 2-D detector pattern contains two peaks which are, from left to right, the [111] and [002] diffraction peaks. The shape of the diffraction pattern is neither a ring (polycrystalline) nor a dot (single crystal) but a broadened line, suggesting textured growth of the $Pt_3Sn$ layer, and the intensity indicates the majority of crystals are ordered along [111] the direction. The crystalline structure of $Pt_3Sn$ along [111] and [002] is shown in the inner set of Figure 2. With a space group of Pm-3m, the [111] direction of $Pt_3Sn$ has a 3-fold symmetry, which is favored for BMER observation[25,27].

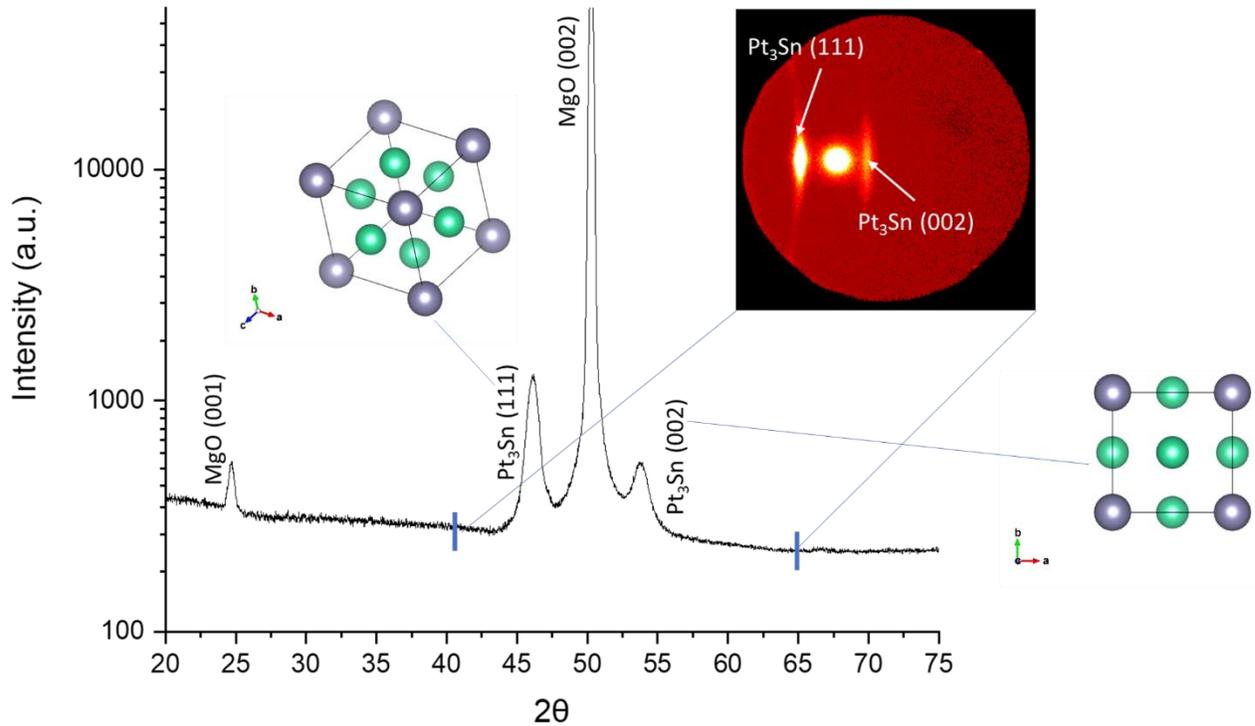

Figure 2. XRD data of the (001)MgO/Pt$_3$Sn(15)/capping sample. The inset shows the original 2D detector image in the range of 2θ containing the [111] and [002] Pt$_3$Sn peaks. The inner set figures show the crystalline structure of Pt$_3$Sn along [111] and [002], respectively.

The single layer Pt$_3$Sn (15nm) film is then patterned into a 10*30 μm$^2$ Hall bar for BMER measurements. An AC current with 2mA amplitude and 133 Hz frequency is applied to the sample, and the second harmonic signal is measured along the longitudinal direction with an SR830 lock-in amplifier, as shown in Figure 3(a). An external field is applied with an angle θ with respect to the current direction in the XY plane. The measured angular-dependent BMER signal under 30000 Oe external field is shown in Figure 3(b). The field dependence of the second harmonic voltage is shown in Figure 3(c), suggesting a linear relation with the field which is an identical feature for BMER. The BMER coefficient $\chi$ is defined with the following equation [25,27]:

$$\chi = \frac{2\Delta R}{RJH} \quad (1),$$

where is $\Delta R$ is the change of resistance and $R$ is the resistance value, $J$ is the current density and $H$ is the applied field.

The calculated BMER coefficient for $Pt_3Sn$ is 0.0067 $nm^2A^{-1}Oe^{-1}$, which is sizable at room temperature compared to previous research [25,39]. $Pt_3Sn$ has shown the existence of bulk topological states, as well as a charge-to-spin conversion ratio of up to 0.4 [38]. These features have contributed to a large BMER value, and the existence of BMER can in turn become the indicator of the existence of topological states and spin-momentum locking. Compared to characterization methods such as ARPES, using BMER for spin-momentum locking characterization may have other advantages besides its fit for polycrystalline samples. Existence of BMER at different temperatures [25,27] and a simple two-terminal measurement setup make it suitable to be integrated into different transport measurement setups.

The BMER signal is also measured in XZ and YZ planes with $\beta$ and $\psi$ angular dependent measurements, as defined in Figure 3(a). Measurement results in the XZ plane shows a similar trend and the same value of BMER compared to that in the XY plane. The BMER signal vanishes in the YZ plane, as shown in Figure 3(d). Compared to previous BMER research, the angular dependence of BMER values in XZ and YZ planes are different in two aspects [25-27]: (1) BMER signal is observed in the YZ plane. (2) The XZ plane signal has exactly the same angular dependence as the XY plane signal.

BMER exists in planes with 3-fold symmetry [25-27]. In previous BMER research in single crystals [25-27], the signal in the XY plane stays the same regardless of the current direction, the signal in the YZ plane has a 120-degree period of current direction dependence, and the signal in XZ plane is the difference of the signals in XY and YZ plane [25,27]. As shown in Figure 2, the $Pt_3Sn$ layer is polycrystalline with most grains along [111] directions. The $Pt_3Sn$ grains are in different directions, but the [111] textured grain has a 3-fold symmetric axis pointing toward z direction. In the XY plane, the grains lie randomly as shown in Figure 3(e), and the direction with the largest positive BMER value is indicated as the triangles in Figure 3(e). As the Hall bar device contains multiple grains, each grain has a BMER contribution that differs in magnitude and direction. Contributions of the YZ plane from different crystallites cancel each other in the Hall bar device, which results in a zero net signal. Similarly, the crystalline direction dependence part in the XZ plane signal is also canceled, resulting in a net angular-dependent signal that is the same as the XY plane signal. The experimental observation of BMER in textured $Pt_3Sn$ demonstrates that BMER can exist in polycrystalline systems with isotropic values regardless of crystalline directions, which can provide strong evidence of the existence of spin-momentum locking in polycrystalline materials.

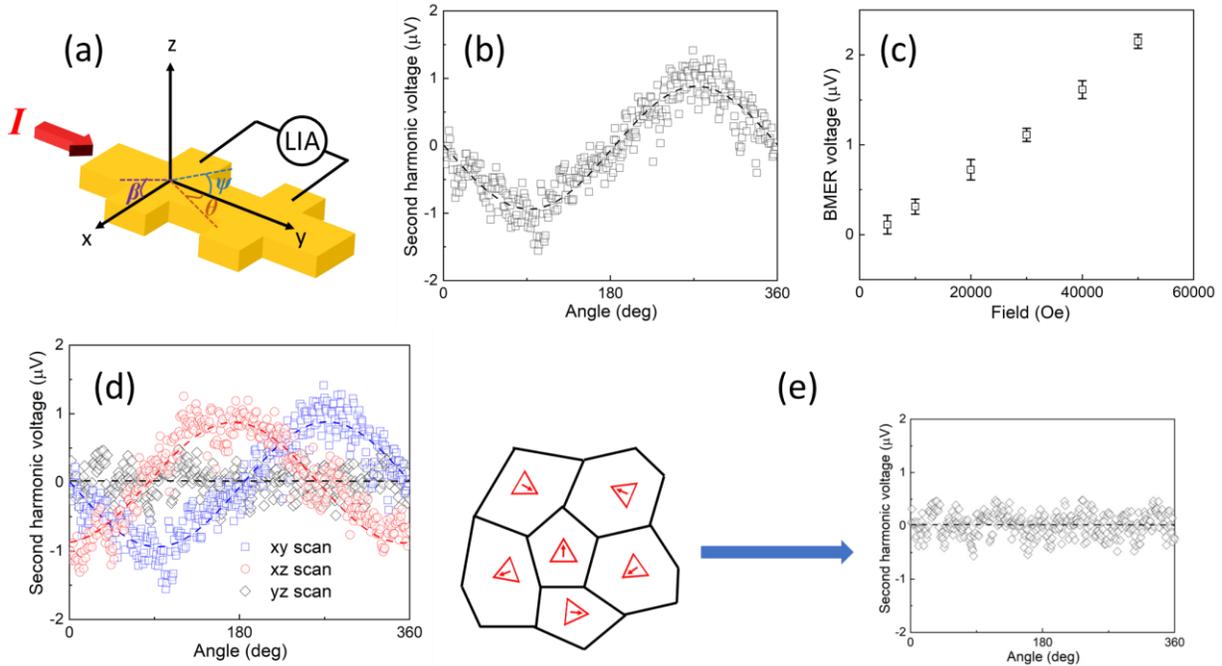

Figure 3. (a) Illustration of the second harmonic measurement and XY($\theta$), YZ($\psi$) and XZ($\beta$) scans. (b) Observed second harmonic voltage of XY scan. (c) Field dependence of BMER voltage shows the voltage is proportional to the applied field. (d) Second harmonic voltage for XY, YZ and XZ scans, (e) Illustration of the absence of the z direction signal in YZ and XZ scan. Crystallites at different directions contribute to z direction signals which cancel with each other and result in a flat line in YZ scan.

To further study this material system, two stack structures of $Pt_3Sn$(10nm)/Al(4nm) (sample 1) and $Pt_3Sn$(10nm)/CoFeB(6nm)/MgO(2nm)/Al(4nm) (sample 2) are deposited, and patterned into the same Hall bar structure as shown in Figure 3(a). The resistance values for the measured device on sample 1 and sample 2 are 730 $\Omega$ and 393 $\Omega$, respectively. The two stack structures are measured under the same voltage (1.03 V), and the measurement current values for sample 1 and sample 2 are 1.4 mA and 2.6 mA, respectively. With the existence of the ferromagnetic layer,

USMR and BMER should exist simultaneously in sample 2. Figure 4(a) shows the observed second harmonic signal for the structure at different field values in sample 2 (divided by 2.6 mA current), where both USMR + Thermal contributions (purple arrow, field independent portion of the signal) and BMER (green arrow, field dependent portion of the signal) are observed. The field dependence of the second harmonic signal is shown in Figure 4(b), where the intercept is the USMR and thermal contributions and the linear, field-dependent part is the BMER.

To qualitatively study the BMER value, current shunting from the CoFeB layer needs to be considered. Another two samples $Pt_3Sn(10nm)/CoFeB(3,4.5nm)/MgO(2nm)/Al(4nm)$ are deposited and patterned together with sample 2 for resistivity measurement. The resistance is measured for this set of samples and is fit with the parallel model: $\frac{1}{R_{total}} = \frac{w\sigma}{l} t_{CFB} + \frac{1}{R_{PtSn}}$, where $w(l)$ is the Hall channel width (length) and $\sigma$ is the conductivity of the CoFeB layer. The resulting resistance contribution for the $Pt_3Sn$ layer in sample 2 is 703 Ω, which is similar to the resistance of sample 1 (730 Ω). The resistance value is then used to calculate the current shunting in sample 2, and the current in layer $Pt_3Sn$ $I_{PtSn}$=1.46 mA is obtained. Since the value of $I_{PtSn}$ is similar to the applied current of sample 1, a direct comparison of $\Delta R/R$ in sample 1 and sample 2 can reveal the BMER coefficient difference in the $Pt_3Sn$ layer. The value $\Delta R/R$ versus the applied field is shown in Figure 4(c), where sample 2 has a BMER value ~2 times larger than sample 1. The BMER coefficient for sample 1 is 0.0059 $nm^2A^{-1}Oe^{-1}$, which is similar to the coefficient (0.0067 $nm^2A^{-1}Oe^{-1}$) in the $Pt_3Sn$(15nm) sample. With an adjacent CoFeB layer, the BMER coefficient changes to 0.0113 $nm^2A^{-1}Oe^{-1}$. This result shows that adding a ferromagnetic layer can lead to an increment of the BMER value in the $Pt_3Sn$ layer.

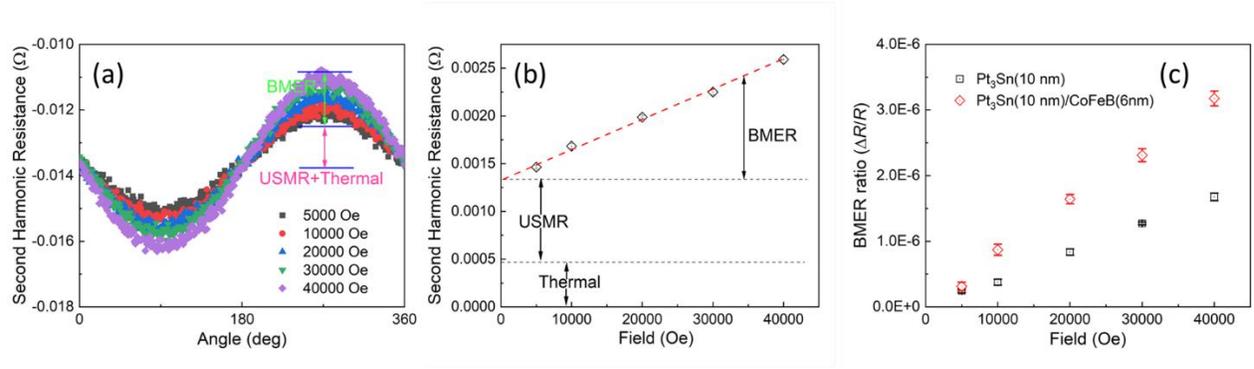

Figure 4 (a) Second harmonic resistance of sample 2 ($Pt_3Sn$(10nm)/CoFeB(6nm)/capping layer), the applied current is 2.6 mA. (b) The field-dependent plot shows the BMER and USMR+Thermal contributions. (c) BMER ratio for $Pt_3Sn$ samples with and without an adjacent CoFeB layer. The adjacent CoFeB layer can increase the BMER value by a factor of 2.

The enhancement of BMER may have multiple origins. According to previous research, the value of BMER can be given by the following equation [25]:

$$R_{BMER} = \frac{El}{w}\left(\frac{36\pi\lambda^2 \varepsilon_F g\mu_B}{ev_F^5 \hbar^4}\right) \quad (2),$$

where $E$ is the electric field, $l$ ($w$) is the length (width) of the device, $\lambda$ is the hexagonal warping term [25,40] and $\varepsilon_F$ ($v_F$) is the Fermi energy (Fermi velocity). Note that an out-of-plane term should also exist, but the textured polycrystalline feature of the $Pt_3Sn$ layer has canceled the contribution of the $\cos3\theta$ dependence as mentioned in the above discussion. Previous research has shown that an adjacent FM layer may open up a gap at the topological band structures, [41,42] which may account for the increment of the BMER value with an adjacent FM layer, as shown in Figure 5. The Fermi level may not lie directly at the Dirac cone, and BMER or spin momentum locking can still exist regardless of the Fermi level location[25,43]. Without opening the gap, the Fermi velocity of the electrons near the Dirac point should be constant, as shown in Figure 5(a). Opening the gap will flatten the bottom part of the conduction band, which will reduce the Fermi velocity $v_F = \frac{dE}{\hbar dk}$, as shown in Figure 5(b). According to equation (2), the Fermi velocity has a power of 5. The topological features such as the spin momentum locking can still exist with a small gap opened [44,45], but even a small change in Fermi velocity can result in a large change in BMER value.

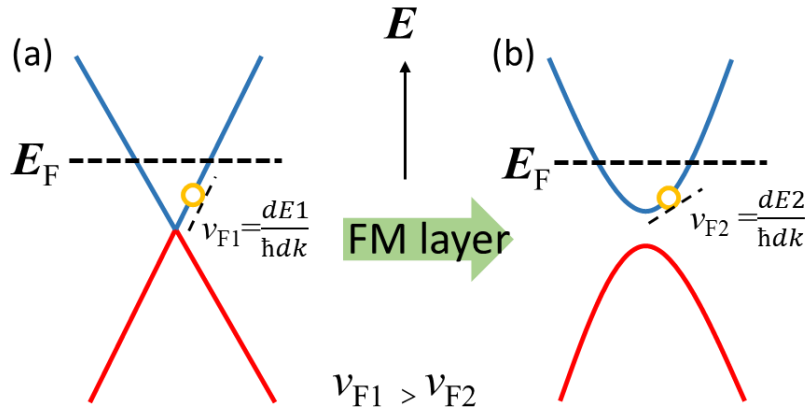

Figure 5. Illustration of the BMER value increment with an adjacent FM layer. The original band structure is shown in (a) where the Fermi velocity is constant. The adjacent layer can open a gap at the Dirac cone as shown in (b), which will reduce the Fermi velocity of the electrons.

Recently, the observation of UMR in antiferromagnetic materials has attracted attention [46]. The large effective magnetic field induced by antiferromagnetic spin canting can enhance the distortion of the Fermi counter, which can enhance the existing field-dependent magnetoresistance signal (including BMER and the magnetoelectric resistance introduced by Rashba splitting [46]) and induce an observable UMR in antiferromagnetic/heavy metal bilayers. The enhancement of the magnetoresistance is induced by the strong spin canting effective field, which does not exist in ferromagnetic systems. Here, we suspect the band opening is the dominant factor for the enhancement of BMER value.

In conclusion, we observed BMER in sputtered $Pt_3Sn$ which has both surface and bulk topological states. The BMER coefficient in a single 15nm $Pt_3Sn$ layer can reach a value of 0.0067 $nm^2A^{-1}Oe^{-1}$. The sputtered $Pt_3Sn$ has a textured polycrystalline structure, which results in the absence of out-of-plane BMER. With an adjacent CoFeB FM layer, the BMER value was enhanced from 0.0059 $nm^2A^{-1}Oe^{-1}$ to 0.0113 $nm^2A^{-1}Oe^{-1}$. We attributed the enhancement of BMER to the band opening induced by the FM layer, which results in a smaller Fermi velocity near the Dirac cone. The observation of the BMER enhancement paves the way for the manipulation of BMER value, and different contributions resulting in a higher BMER value can be further explored.

Acknowledgments:

This work was supported, in part, by SMART, one of the seven centers of nCORE, a Semiconductor Research Corporation program, sponsored by the National Institute of Standards and Technology (NIST) and by the UMN MRSEC program under Award No. DMR-2011401. This work utilized the College of Science and Engineering (CSE) Characterization Facility at the University of Minnesota (UMN) supported, in part, by the NSF through the UMN MRSEC program. Portions of this work were conducted in the Minnesota Nano Center, which is supported by the National Science Foundation through the National Nano Coordinated Infrastructure Network (NNCI) under Award Number ECCS-2025124. D. Zhang and T. Peterson are partially supported by ASCENT, one of six centers of JUMP, a Semiconductor Research Corporation program that is sponsored by MARCO and DARPA. J.-P.W. also acknowledges support from Robert Hartmann Endowed Chair Professorship.

The data that support the findings of this study are available from the corresponding author upon reasonable request.